\documentclass[conference]{IEEEtran}
\IEEEoverridecommandlockouts
\usepackage{cite}
\usepackage{amsmath,amssymb,amsfonts}
\usepackage{algorithmic}
\usepackage{graphicx}
\usepackage{textcomp}
\usepackage{xcolor}
\def\BibTeX{{\rm B\kern-.05em{\sc i\kern-.025em b}\kern-.08em
    T\kern-.1667em\lower.7ex\hbox{E}\kern-.125emX}}
\begin{document}

\title{AR-Assisted Surgical Care via 5G networks for First Aid Responders
\thanks{This work is partially funded by the 5G-Epicentre Horizon 2020 project.}
}

\author{\IEEEauthorblockN{Manos Kamarianakis}
\IEEEauthorblockA{\textit{University of Crete, FORTH} \\
\textit{ORamaVR SA}\\
Heraklion, Crete \\
manos.kamarianakis@oramavr.com}
\and
\IEEEauthorblockN{Antonis Protopsaltis}
\IEEEauthorblockA{\textit{University of Western Macedonia} \\
\textit{ORamaVR SA}\\
Kozani, Greece \\
antonis.protopsaltis@oramavr.com}
\and
\IEEEauthorblockN{George Papagiannakis}
\IEEEauthorblockA{\textit{University of Crete, FORTH} \\
\textit{ORamaVR SA}\\
Heraklion, Greece \\
george.papagiannakis@oramavr.com}
}

\maketitle

\begin{abstract}

Surgeons should play a central role in disaster planning and
management due to the overwhelming number of bodily injuries that are
typically involved during most forms of disaster. In fact, various
types of surgical procedures are performed by emergency medical teams
after sudden-onset disasters, such as soft tissue wounds, orthopaedic
traumas, abdominal surgeries, etc \cite{coventry2019surgical,Birrenbach2021jmir}. 
HMD-based Augmented Reality (AR), using state-of-the-art hardware 
such as the Magic Leap or the
Microsoft HoloLens, have long been foreseen as a key enabler for
clinicians in surgical use cases \cite{zikas2022virtual}, especially for procedures performed
outside of the operating room. In such condtions, monolithic HMD applications fail to maintain important factors such as user-mobility, battery life, and Quality of Experience (QoE), leading to considering a distributed cloud/edge software architecture. Toward this end, 5G and cloud computing will be a central
component in accelerating the process of remote rendering computations and image transfers to wearable AR devices.

This  paper  describes  the  Use  Case (UC) ”AR-assisted emergency
surgical care”,  identified  in the context of the 5G-EPICENTRE  
EU-funded  project. Specifically, the UC will experiment with holographic AR
technology for emergency medical surgery teams, by overlaying
deformable medical models directly on top of the patient body parts,
effectively enabling surgeons to see inside (visualizing bones, blood
vessels, etc.) and perform surgical actions following step-by-step
instructions. The goal is to combine the computational and
data-intensive nature of AR and Computer Vision algorithms with
upcoming 5G network architectures deployed for edge computing so 
as to satisfy real-time interaction requirements and provide an 
efficient and powerful platform for the pervasive promotion of  
such applications. Toward this end, the authors have 
adapted the psychomotor Virtual Reality (VR) surgical training 
solution, namely MAGES \cite{papagiannakis2020mages,papagiannakis2018transforming}, developed by the 
ORamaVR company. By developing the necessary Virtual Network Functions 
(VNFs) to manage data-intensive services (e.g., prerendering, 
caching, compression) and by exploiting available network resources 
and Multi-access Edge Computing (MEC) support, provided by the
5G-EPICENTRE infrastructure, this UC aims to provide powerful AR-based tools, usable on site, to first-aid responders.
\end{abstract}


\section{Introduction}
\begin{figure}
    \centering
    \includegraphics{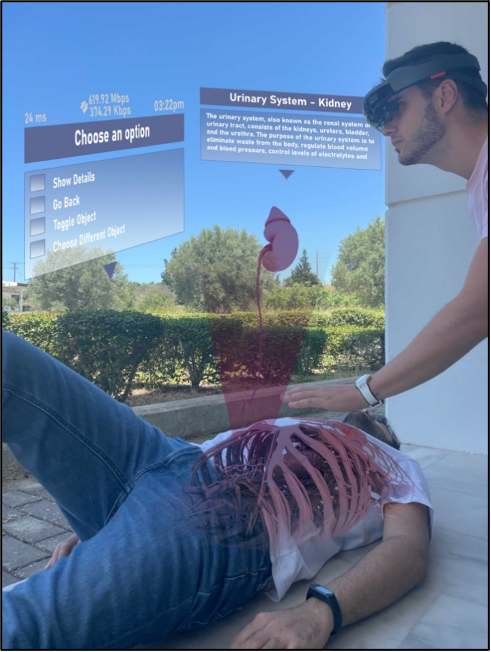}
    \caption{The PPDR responder uses an AR HMD to see overlayed info and deformable objects on top of the patient. Envisioned example of UI layout.}
    \label{fig:5g_sample}
\end{figure}

Main purpose of this UC is to perform various experimentations on the 
5G infrastructure via an AR-assisted surgical care application for Public 
Protection and Disaster Relief (PPDR). Ultimately, the UC aims to provide 
first-aid responders, situated on a disaster site, with a powerful tool that will help 
save lives in peril. Using untethered highly-portable AR head-mounted displays (HMDs), 
the PPDR responders will be able to visualise various deformable internal body parts 
on top of the patient as if they had super-human vision. 
Furthermore, an instruction set will allow them to perform various operations 
with the possibility of remote guidance from a medical expert 
(see Fig.~\ref{fig:5g_sample}). 
The current state-of-the-art, driven, among others, by ORamaVR’s innovative 
VR Engine \emph{MAGES}, allows the usage of tethered and untethered VR HMDs to perform 
similar tasks such as VR surgeries. 
In the particular setup, VR HMDs have to either be connected with a cable
to a VR-ready local computer (tethered device which provides high performance GPU/CPU but limited user-mobility), or run standalone (untethered device provides user-mobility, but relatively decreased GPU/CPU performance), to create a high 
Frames-per-Second (FPS) VR scene. Within the 5G-EPICENTRE project, the power of 
MAGES is leveraged to untethered AR HMDs. By offloading the processor-heavy 
rendering operations to the edge/cloud services
provided by the 5G-EPICENTRE infrastructure, we lighten the processing burden of AR HMDs 
to mostly receive and project the rendered scenes and only broadcast the user interactions 
back to the edge via a 5G connection. Since the heavy processing will not be 
performed on the device, a significant decrease in the HMD's energy consumption is expected while the use of inexpensive, light-weight HMDs now becomes a 
possibility. As the PPDR first-aid responder may be located outdoors at any incident site, 
the increased energy autonomy of the portable device will reduce environmental impact 
and enable greater portability of the whole project.

\section{Benefits from Transitioning to Cloud/Edge Resources} 
\label{sec:benefits_from_transitioning_to_cloud_edge_resources}

By exploiting CPU and GPU resources that are available on the Cloud/Edge continuum, 
we gain numerous advantages in comparison to the classic setup, where a 
monolithic application component is installed and run on (un)tethered HMD. 

Indeed, in the classic tethered HMD setup, all processes involving storage, rendering and 
compression are run on a VR-ready PC and therefore we have a serious limitation in terms of 
mobility/portability. The utilization of untethered HMDs partially lifts the mobility constraint, but with two drawbacks: a) such devices have inadequate hardware specifications, in terms of CPU power (ARM-based processors) and of GPU power (limited RAM, no CUDA enabled), and b) limited battery life, as they execute all processor-heavy rendering processes, that drain battery fast. Finally, 
in classic multi-user setups, the user initializing a session is  considered the host of the
multi-user session, also referred as \emph{master server}. 
The QoE of the multi-user session depends on the master server's hardware and bandwidth resources, which is 
an additional point to consider when multiple users are involved.

On the other hand, the 5G-EPICENTRE platform allows many of these constraints to 
be lifted within the context of a cloud/edge based application. Firstly, the platform 
will provide powerful cloud-edge resources to the AR PPDR application, transforming the monolithic application to a Software-as-a-Service, which will be available for consumption even 
for untethered, affordable HMDs of lower specifications. 
By lifting the hardware restrictions, we can a) design a device agnostic framework and b)
use lightweight HMDs to achieve high-mobility and increased battery life, which is essential for PPDR missions. Lastly, the deployment of the netapp in cloud/edge resources guarantees flawless 
multi-user sessions  as they will no longer depend on the network characteristics of the first user.


\section{Expected Results}

The design of an AR-Assisted Surgical Care application that exploits resources 
on the cloud-edge continuum, is a difficult task that, when completed, will yield 
satisfying results. In the envisioned setup, the heavy processes such as Physics and 
Scene Rendering, will be performed on an Edge Node of high CPU/GPU performance. 
The HMD's resources will be used mainly to send user input (movement and triggers) to the edge netapp and decompress/project the stream. In this way, we expect to achieve increased mobility 
for existing HMDs, but also enable the usage of low-spec HMDs. The rendered scene
will be streamed compressed (e.g., WebRTC) to the HMD via the 5G network, which 
allows Low latency and  high bandwidth. In this way, we may stream high quality and fidelity images, without compromising QoE; as long as a minimal latency 
is kept, user immersion in AR can be preserved. In case of network characteristics fluctuations, QoE can still be guaranteed by exploiting adaptive resolution 
techniques. Specifically, we may choose to stream lower-quality images whenever the 
5G-EPICENTRE or application analytics engine detects a bottleneck in the sequence. 

To be more specific, the key performance indicators (KPIs) that the UC aims to 
achieve within the 5G-EPICENTRE project are the following: a) Round Trip Time (RTT) less than 
$7$ms, b) total aggregated bandwidth of more than $0.7$ Gb/s and c) a battery life increase
of more than $30$\% for untethered AR HMDs, compared to the classic setup. 

\section{Deployment of UC Components}
\label{sec:components}
\begin{figure}
    \centering
    \includegraphics[width=0.5\textwidth]{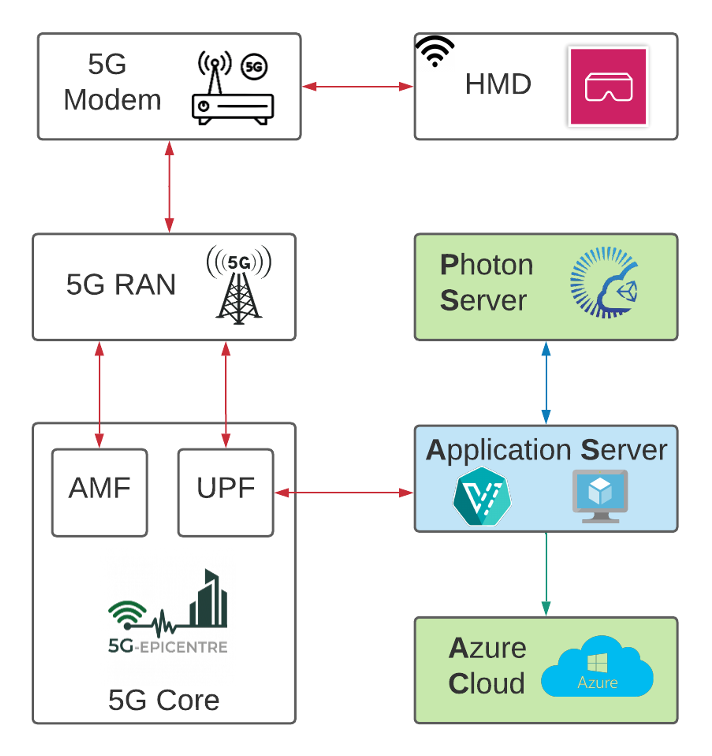}
    \caption{Envisioned layout of the UC components (see \ref{sec:components}). 
    Blue and Green nodes correspond to edge and cloud resources respectively.}
    \label{fig:5g_diagram}
\end{figure}

The components that are used for this UC, are depicted in Fig.~\ref{fig:5g_diagram}. 
Specifically, the components developed by the ORamaVR team are a) the application 
running a) on the HMD and b) the application server containerized via kubevirt, on an 
edge node (depicted in blue). The latter application 
exploits cloud-based services such as the Azure Cloud for database management and the 
Photon Server for multi-user sessions (depicted in green). The the application server instantiation on the edge is performed via the 5G-EPICENTRE platform that 
handles all necessary intermediate tasks such as optimal edge node discovery/availability, 
network slicing, deployment sequence, handshaking, etc. Furthermore, the platform 
also monitors and logs useful network metrics throughout the use of the AR application
and identifies, in real-time, potential bottlenecks that in turn activate adaptive 
rendering mechanisms. The connection of the AR HMD application to the 5G network and 
hence the 5G-EPICENTRE platform is accomplished via a 5G modem, to compensate the lack
of 5G connection ports in most AR HMDs.

\section{Experimentation}

The UC experimentation on the 5G-EPICENTRE platform is expected to be performed in the following phases:
\begin{enumerate}
\item Identification of the capabilities of the 5G network architectures and services provided by the testbed provider as well as the edge computing network services.
\item Implementation of a suitable netapp version of an AR application, which will run on the edge.
\item Development of the necessary VNFs to manage data-intensive services (e.g., prerendering, caching, compression), exploiting available network resources and multi-access edge computing (MEC) support provided by 5G-EPICENTRE.
\item Determine potential performance bottlenecks of operating a monolithic AR application. 
\item Suggest offloading of the software architecture through microservices without adding excessive inter-services latency. 
\item Fine tuning of the AR application to fully exploit available network resources and MEC support provided by 5G-EPICENTRE.
\item Fulfilment of the KPIs regarding bandwidth usage and latency, as well as device energy consumption.
\item Compilation of a list of insights towards a better understanding of the 5G network utilisation in conjunction with state-of-the-art AR applications and present the tools that were used throughout the experimentation.
\end{enumerate}

To demonstrate the added value of the experiment, various test cases have been specified:
\begin{itemize}
\item Measurement of end-to-end latency using different configurations of data compression.
\item Conduct load testing to determine the optimal number of users that the network allows within the latency threshold.
\item Conduct stress testing to determine the number of users that the network can accommodate without inducing network congestion while preserving QoE. 
\item Experimentation with state-of-the-art object recognition methodologies to provide 3D model optimal overlay.
\item Evaluation of different user interfaces in search of the optimal solution that provides a smooth user experience while visualising the proper amount of information streamed from the edge.
\end{itemize}

The deployment and experimentation will take place in Barcelona, from 
June 2022 to June 2023, where the high CPU/GPU node capabilities of the CTTC testbed 
will be exploited. The lab experimentation will focus on RTT, Bandwidth Throughput and
Package Loss. As already mentioned, all tests aim to stress-test the capabilities of the 
5G-EPICENTRE platform. The ultimate goal is to draw QoE-related conclusions from 
the experimentation results and fine-tune various network settings that will be 
reported by the project for future reference.

\section{Acknowledgment} 
\label{sec:ackwnoledgment}
The work presented is funded by the 5G-EPICENTRE Horizon 2020 EU-funded project.

\bibliographystyle{IEEEtran}
\bibliography{IEEEabrv,bibliography}

\end{document}